\documentclass[conference]{IEEEtran}

\usepackage[cmex10]{amsmath}
\usepackage{cite}

\usepackage{latexsym}
\usepackage{graphics}
\usepackage{amssymb}
\usepackage{amsthm}

\usepackage{cite}
\usepackage{array}
\usepackage{graphicx}
\usepackage[linesnumbered,ruled,norelsize,vlined]{algorithm2e}

\usepackage{url}

\def\vectorize{\mathrm{vec}}

\def\kron{\otimes}
\def\tr{\mathrm{tr}}

\def\vectorize{\mathrm{vec}}
\newcommand{\MMSE}[1]{\widehat{#1}_{\mathrm{MMSE}}}
\newcommand{\PEACH}[1]{\widehat{#1}_{\mathrm{PEACH}}}
\newcommand{\WPEACH}[1]{\widehat{#1}_{\mathrm{W}\textrm{-}\mathrm{PEACH}}}
\newcommand{\MVU}[1]{\widehat{#1}_{\mathrm{MVU}}}

\newcommand{\vect}[1]{\mathbf{#1}}

\theoremstyle{plain}
\newtheorem{remark}{Remark}
\newtheorem{theorem}{Theorem}

\newtheorem{proposition}{Proposition}
\newtheorem{lemma}{Lemma}

\IEEEoverridecommandlockouts

\begin{document}

\title{Low-Complexity Channel Estimation in\\ Large-Scale MIMO using Polynomial Expansion}

\author{\IEEEauthorblockN{Nafiseh Shariati\IEEEauthorrefmark{1}, Emil~Bj\"ornson\IEEEauthorrefmark{1}\IEEEauthorrefmark{3}, Mats~Bengtsson\IEEEauthorrefmark{1}, and M\'erouane~Debbah\IEEEauthorrefmark{3} \thanks{E.~Bj\"ornson is funded by the International Postdoc Grant 2012-228 from the Swedish Research Council. This research has been supported by the ERC Starting Grant 305123 MORE (Advanced Mathematical Tools for Complex Network Engineering).}}
\IEEEauthorblockA{\IEEEauthorrefmark{1}Signal Processing Lab, ACCESS Linnaeus Centre, KTH Royal Institute of Technology, Stockholm, Sweden}
\IEEEauthorblockA{\IEEEauthorrefmark{3}Alcatel-Lucent Chair on Flexible Radio, SUPELEC, Gif-sur-Yvette, France}
E-mail: \{nafiseh, mats.bengtsson\}ee.kth.se, \{emil.bjornson, merouane.debbah\}@supelec.fr}
\maketitle

\maketitle

\begin{abstract}

This paper considers pilot-based channel estimation in large-scale multiple-input multiple-output (MIMO) communication systems, also known as ``massive MIMO''. Unlike previous works on this topic, which mainly considered the impact of inter-cell disturbance due to pilot reuse (so-called \emph{pilot contamination}), we are concerned with the computational complexity. The conventional \emph{minimum mean square error (MMSE)} and \emph{minimum variance unbiased (MVU)} channel estimators rely on inverting covariance matrices, which has cubic complexity in the multiplication of number of antennas at each side. Since this is extremely expensive when there are hundreds of antennas, we propose to approximate the inversion by an $L$-order matrix polynomial. A set of low-complexity Bayesian channel estimators, coined Polynomial ExpAnsion CHannel (PEACH) estimators, are introduced. The coefficients of the polynomials are optimized to yield small mean square error (MSE). We show numerically that near-optimal performance is achieved with low polynomial orders. In practice, the order $L$ can be selected to balance between complexity and MSE. Interestingly, pilot contamination is beneficial to the PEACH estimators in the sense that smaller $L$ can be used to achieve near-optimal MSEs.
\end{abstract}

\IEEEpeerreviewmaketitle

\section{Introduction}

 MIMO techniques can bring substantial improvements in spectral efficiency to wireless systems, by increasing the spatial reuse.
While $8 \times 8$ MIMO transmission has found its way into standards such as LTE-Advanced \cite{Holma2012a}, there is an increasing interest in equipping base stations with much larger
antenna arrays \cite{Jose2011b,Hoydis2013a,Rusek2013a}. Such large-scale MIMO, or ``massive MIMO'', techniques can give unprecedented spatial resolution, enabling a very dense spatial reuse that potentially can keep up with the rapidly increasing demand for wireless connectivity.

A major limiting factor in large-scale MIMO is the availability of accurate channel state information (CSI). This is since high spatial resolution can only be exploited if the propagation environment is precisely known. CSI is typically acquired by sending pilot signals and estimating the channel coefficients from the received signals \cite{Yin2013a,Mueller2013a,Kay1993a,Kotecha2004a,Liu2007a,Bjornson2010a}.
The Bayesian MMSE estimator can be applied \cite{Kay1993a,Kotecha2004a,Liu2007a,Bjornson2010a} if the channel statistics are known, while the MVU estimator is applied otherwise \cite{Kay1993a}.

These channel estimators basically solve a linear system of equations, or equivalently multiply the received signal with an inverse of the covariance matrices, which is a mathematical operation with cubic computational complexity. Therefore, it is
very computationally expensive to compute the MMSE and MVU estimates in large-scale MIMO systems. The high complexity can be avoided under ideal propagation conditions where all covariance matrices are diagonal, but large-scale MIMO channels typically have a distinct spatial correlation due to insufficient antenna spacing and richness of the propagation environment \cite{Rusek2013a}. Moreover, the necessary pilot reuse in cellular networks creates spatially correlated inter-cell interference, known as \emph{pilot contamination}, that reduces the estimation performance and the spectral efficiency \cite{Jose2011b,Hoydis2013a,Rusek2013a,Yin2013a,Mueller2013a}.

Similar complexity issues appear in multiuser detection, where both the decorrelating detector and the linear MMSE detector involve matrix inversions \cite{Moshavi1996a}.
A common low-complexity approach is reduced-rank filtering \cite{Honig2001a}. This can be achieved by \emph{polynomial expansion (PE)}, where the matrix inverse is approximated by an $L$-order matrix polynomial \cite{Moshavi1996a,Honig2001a,Sessler2005a,Hoydis2011d}. PE-based detectors are versatile since the structure enables simple multistage hardware implementation \cite{Moshavi1996a} and the order $L$ needs not to scale with the system size to achieve near-optimal performance \cite{Honig2001a}. Therefore, $L$ is simply selected to balance between complexity and detection performance. A main problem is to select the coefficients of the polynomial to achieve high performance at small $L$; the optimal weights are expensive to compute \cite{Moshavi1996a}, but alternatives based on appropriate scalings \cite{Sessler2005a} and asymptotic analysis \cite{Hoydis2011d} exist.

Inspired by the prior works in detection, in this paper, we propose a set of low-complexity channel estimators that we call \emph{Polynomial ExpAnsion CHannel (PEACH)} estimators. These novel estimators approximate the MMSE estimator by replacing the matrix inversion with a polynomial expansion. The coefficients of the polynomial are optimized to yield minimal MSE at any fixed polynomial order $L$, while keeping the low complexity. The PEACH estimators are evaluated under different propagation/interference conditions and show remarkably good performance at low polynomial orders.

\textbf{Notation:}
Boldface (lower case) is used for column vectors, $\vect{x}$, and
(upper case) for matrices, $\vect{X}$. Let  $\vect{X}^T$,
$\vect{X}^H$, and $\vect{X}^{-1}$ denote the transpose, the conjugate
transpose, and inverse of $\vect{X}$, respectively. The Kronecker product of $\vect{X}$ and $\vect{Y}$ is
denoted $\vect{X} \otimes \vect{Y}$, $\vectorize(\vect{X})$ is the
 vector obtained by stacking the columns of $\vect{X}$,
$\tr( \vect{X} )$ denotes the trace, and $\|\vect{X}\|_F$ is the Frobenius norm. The notation $\triangleq$ denote definitions, while
$\mathcal{O}(M^x)$ means that the complexity is bounded by $ C M^x$ for some $C>0$.

\section{Problem Formulation}

\begin{figure}
   \begin{center}
   \includegraphics[width=.9\columnwidth]{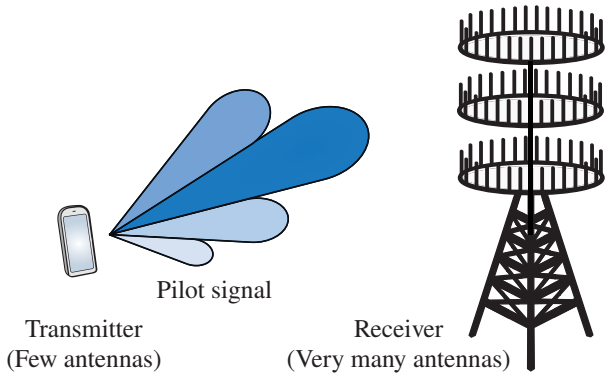}\\
 \caption{Illustration of pilot signaling in a large-scale $N_t \times N_r$ MIMO system, where $N_r \gg N_t$. The complexity of conventional channel estimators is very large in these systems, which calls for low-complexity alternatives.}
   \label{fig:problem_illustration}
   \end{center}
\end{figure}

This paper revisits the problem of estimating a quasi-static flat-fading MIMO channel $\vect{H} \in \mathbb{C}^{N_r \times N_t}$, where $N_t$ is the number of transmit antennas and $N_r$ is the number of receive antennas. Similar to \cite{Kotecha2004a,Liu2007a,Bjornson2010a}, the transmitter sends a fixed predefined pilot matrix $\vect{P} \in \mathbb{C}^{N_t \times B}$ over $B \geq 1$ symbol times; see Fig.~\ref{fig:problem_illustration}. The receiver tries to obtain $\vect{H}$ from the received signal $\vect{Y} \in \mathbb{C}^{N_r \times B}$, given by
\begin{equation} \label{eq_trainingmodel}
\vect{Y} = \vect{H} \vect{P} + \vect{N}
\end{equation}
where the disturbance $\vect{N} \in \mathbb{C}^{N_r \times B}$ is assumed to be circularly-symmetric complex Gaussian noise: $\vectorize(\vect{N}) \sim \mathcal{CN}(\vect{0},\vect{S})$. The  disturbance covariance matrix $\vect{S} \in \mathbb{C}^{N_r B \times N_r B }$ is positive definite and can include both regular uncorrelated receiver noise and different types of interference from other systems. The analysis herein holds for any $\vect{S}$, but some typical special cases will be considered in Section \ref{subsection:pilot-contamination}.

The channel matrix is modeled as Rayleigh fading with $\vectorize(\vect{H}) \sim \mathcal{CN}(\vect{0},\vect{R})$, where the channel covariance matrix $\vect{R} \in \mathbb{C}^{N_t N_r \times N_t N_r}$ is positive semi-definite. Observe that $\vect{R}$ is generally \emph{not} a scaled identity matrix, but describes the spatial propagation environment. The matrices $\vect{R}$ and $\vect{S}$ are assumed known at the transmitter and receiver.

If the channel statistics are known at the receiver, the Bayesian MMSE estimator of the MIMO channel is \cite{Kay1993a,Kotecha2004a,Liu2007a,Bjornson2010a}
\begin{equation} \label{eq_MMSE_channel_matrix}
\begin{split}
\vectorize(\MMSE{\vect{H}}) = \vect{R} \widetilde{\vect{P}}^H
\left(\widetilde{\vect{P}} \vect{R} \widetilde{\vect{P}}^H +
\vect{S} \right)^{-1}  \vectorize(\vect{Y})
\end{split}
\end{equation}
where $\widetilde{\vect{P}} \triangleq (\vect{P}^T \! \kron \, \vect{I})$. The performance is measured in $\textrm{MSE} =\mathbb{E}\{ \|\vect{H} - \MMSE{\vect{H}} \|_F^2 \} = \tr \left( ( \vect{R}^{-1} + \widetilde{\vect{P}}^H
\vect{S}^{-1} \widetilde{\vect{P}}
 )^{-1} \right)$.

Alternatively, if the channel distribution is unknown to the receiver, the classic MVU estimator is \cite[Chapter 4]{Kay1993a}
\begin{equation} \label{eq_MVU_channel_matrix}
\begin{split}
\vectorize(\MVU{\vect{H}}) = \left(\widetilde{\vect{P}}^H \vect{S}^{-1} \widetilde{\vect{P}} \right)^{-1} \widetilde{\vect{P}}^H \vect{S}^{-1} \vectorize(\vect{Y}).
\end{split}
\end{equation}
The performance measure is then the estimation variance $\mathbb{E}\{ \|\vect{H} - \MVU{\vect{H}} \|_F^2 \} = \tr \left( ( \widetilde{\vect{P}}^H
\vect{S}^{-1} \widetilde{\vect{P}} )^{-1} \right)$. Obviously,
\begin{equation}
\tr \left( ( \vect{R}^{-1} + \widetilde{\vect{P}}^H
\vect{S}^{-1} \widetilde{\vect{P}}
 )^{-1} \right) < \tr \left( ( \widetilde{\vect{P}}^H
\vect{S}^{-1} \widetilde{\vect{P}} )^{-1} \right)
\end{equation}
for any $\vect{R} \neq \vect{0}$, thus the MMSE estimator achieves a better average estimation performance than the MVU estimator since it utilizes the channel statistics.

\subsection{Complexity Issues in Large-Scale MIMO Systems}

The main computational complexity when computing the MMSE and MVU estimators in \eqref{eq_MMSE_channel_matrix} and \eqref{eq_MVU_channel_matrix} lies in solving a linear system of equations or, equivalently, in computing the matrix inversions directly.\footnote{If $\vect{R},\vect{S},\vect{P}$ are all diagonal matrices, the complexity can be greatly reduced by simply inverting the diagonal elements. However, such special cases are of limited practical interest, particularly in large-scale MIMO systems which are prone to non-negligible spatial channel correlation and pilot contamination.} Both approaches have computational complexities that scale as $\mathcal{O}(M^3)$ where $M \triangleq N_t N_r$.\footnote{Note that $\mathcal{O}(M^3)$ refers to the complexity scaling of the classical Gaussian elimination algorithm. The complexity is reduced to $\mathcal{O}(M^{2.8074})$ by the more sophisticated Strassen's algorithm. The exponent can be further reduced, see e.g.~\cite{Williams2012a}, but mainly for academic purposes since extremely large matrices are required to actually benefit from such improved asymptotic behaviors.} This complexity is relatively modest in classic MIMO systems where $2 \times 2$, $4 \times 4$, or $8 \times 8$ are typical configurations.

Recently, there is an increasing interest in large-scale MIMO systems where there might be hundreds of antennas at one side of the link \cite{Jose2011b,Hoydis2013a,Rusek2013a}. To excite all channel dimensions, the pilot length $B$ should be of the same order as $N_t$. Large-scale systems are therefore envisioned to exploit channel reciprocity to always have $N_t < N_r$ in the channel estimation phase---$N_r$ can even be orders of magnitude larger than $N_t$ without degrading the estimation performance \emph{per} antenna element.

Observe that in a potential future large-scale MIMO system with $N_r=200$ and $N_t=20$, the MMSE and MVU estimators would require inverting matrices of size $4000 \times 4000$ (or similarly, solving a linear system of equations with $4000$ unknown variables). This massive operation needs to be redone quite often since $\vect{R}$ and $\vect{S}$ change with time. The purpose of this paper is to develop alternative channel estimators that allow for balancing between computational/hardware complexity and estimation performance.

\begin{remark}
While having Gaussian channels and disturbance is a well-accepted assumption in conventional MIMO systems, the channel modeling for large-scale MIMO is still in its infancy. By increasing the number of antennas we will improve the spatial resolution of the array which eventually may invalidate the rich-scattering assumption that is behind the use of Gaussian channel distributions \cite{Rusek2013a}. However, we stress that the estimators in this paper can be applied and give reasonable performance under any channel and disturbance distributions; this is since \eqref{eq_MMSE_channel_matrix} is also the linear MMSE estimator and \eqref{eq_MVU_channel_matrix} is the best linear unbiased estimator (BLUE) in cases when only the first two moments of $\vect{H}$ and/or $\vect{N}$ are known \cite{Kay1993a,Bjornson2010a}.
\end{remark}

\section{Low-Complexity Bayesian PEACH Estimators}

In this section, we propose several low-complexity Bayesian channel estimators based on the concept of polynomial expansion. To understand the main idea, we first state the following lemma which is easily proved by using standard Taylor series.

\begin{lemma} \label{lemma:inversion-expansion}
For any Hermitian matrix $\vect{X} \in \mathbb{C}^{N \times N}$, with bounded eigenvalues $| \lambda_n(\vect{X})|<1$ for all $n$, it holds that
\begin{equation} \label{eq:inversion-expansion}
\left( \vect{I} - \vect{X} \right)^{-1} = \sum_{l=0}^{\infty} \vect{X}^{l}.
\end{equation}
\end{lemma}

Observe that the impact of $\vect{X}^{l}$ in \eqref{eq:inversion-expansion} reduces with $l$ (as $\lambda_n(\vect{X})^{l}$ for each eigenvalue). It therefore makes sense to consider $L$-order polynomial expansions of the matrix inverse using only the terms $l=0,\ldots,L$. In principle, the inverse of each eigenvalue is then approximated by an $L$-order Taylor polynomial, thus $L$ needs \emph{not} scale with the matrix dimension to achieve a certain accuracy per element. Instead, $L$ can be selected to balance low approximation error with complexity.

In order to apply Lemma \ref{lemma:inversion-expansion} on matrices with any eigenvalue structure, we obtain the next result
which is similar to \cite{Sessler2005a}.

\begin{proposition} \label{prop:inversion-expansion}
For any positive-definite Hermitian matrix $\vect{X}$,
\begin{equation}
\vect{X}^{-1} = \alpha \big( \vect{I} - (\vect{I} - \alpha \vect{X}) \big)^{-1} \approx \alpha \sum_{l=0}^{L} (\vect{I} - \alpha \vect{X})^l
\end{equation}
where the approximation holds with equality when $L \rightarrow \infty$ if $\alpha$ is selected to satisfy $0 < \alpha < \frac{2}{\max_n \lambda_n(\vect{X})}$.
\end{proposition}

\subsection{Unweighted PEACH Estimator}

Applying the approximation in Proposition \ref{prop:inversion-expansion} on the MMSE estimator in \eqref{eq_MMSE_channel_matrix} gives the low-complexity $L$-order \emph{Polynomial ExpAnsion CHannel (PEACH)} estimator
\begin{equation} \label{eq_PEACH_estimator}
\begin{split}
\vectorize(\PEACH{\vect{H}}) = \vect{R} \widetilde{\vect{P}}^H
\sum_{l=0}^{L}  \alpha \big(\vect{I} - \alpha (\widetilde{\vect{P}} \vect{R} \widetilde{\vect{P}}^H + \vect{S} ) \big)^{l}   \vectorize(\vect{Y})
\end{split}
\end{equation}
which does not involve any channel inversions. The computational complexity of \eqref{eq_PEACH_estimator} is $\mathcal{O}(L M^2)$ where $M \triangleq N_t N_r$. Whenever $L \ll M$, this is a large complexity reduction as compared to $\mathcal{O}(M^3)$ for the original MMSE estimator.

\begin{theorem}
The PEACH estimator in \eqref{eq_PEACH_estimator} achieves the MSE
\begin{equation}
\tr \left(\vect{R} +  \vect{R} \widetilde{\vect{P}}^H  \vect{A}_L (\widetilde{\vect{P}} \vect{R} \widetilde{\vect{P}}^H + \vect{S} )  \vect{A}_L^H  \widetilde{\vect{P}} \vect{R} - 2  \vect{R} \widetilde{\vect{P}}^H  \vect{A}_L  \widetilde{\vect{P}} \vect{R} \right)
\end{equation}
where $\vect{A}_L = \sum_{l=0}^{L}  \alpha \big(\vect{I} - \alpha (\widetilde{\vect{P}} \vect{R} \widetilde{\vect{P}}^H + \vect{S} ) \big)^{l}$.
\end{theorem}
\begin{IEEEproof}
Follows from direct computation of the MSE.
\end{IEEEproof}

The PEACH estimator lends itself to efficient multistage hardware implementation where $( \cdot )^{l+1}   \vectorize(\vect{Y})$ is computed from $( \cdot  )^{l}   \vectorize(\vect{Y})$ by multiplying the latter vector with $\big(\vect{I} - \alpha (\widetilde{\vect{P}} \vect{R} \widetilde{\vect{P}}^H + \vect{S} ) \big)$. The hardware  can be similar to the multistage detection implementation illustrated in \cite[Fig.~1]{Moshavi1996a}.

It remains to select the scaling parameter $\alpha$ to satisfy the condition in Proposition \ref{prop:inversion-expansion}.
From a complexity point of view, we easily can select $\alpha$ to be equal to $\frac{2}{\tr(\widetilde{\vect{P}} \vect{R} \widetilde{\vect{P}}^H + \vect{S})}$. On the other hand, we are also interested in choosing $\alpha$ to achieve fast convergence in the polynomial expansion. Among the values that satisfy the condition in Proposition \ref{prop:inversion-expansion}, the choice
\begin{equation}
\begin{split}
\alpha_{\mathrm{PEACH}} = \frac{2}{\max_n \lambda_n(\widetilde{\vect{P}} \vect{R} \widetilde{\vect{P}}^H + \vect{S}) + \min_n \lambda_n(\widetilde{\vect{P}} \vect{R} \widetilde{\vect{P}}^H + \vect{S}) }
\end{split}
\end{equation}
will maximize the asymptotic convergence since the largest and smallest eigenvalue of $\big(\vect{I} - \alpha (\widetilde{\vect{P}} \vect{R} \widetilde{\vect{P}}^H + \vect{S} ) \big)$ become symmetric around the origin \cite{Sessler2005a}. Although the computation of the extreme eigenvalues is generally quite expensive, we note that these can be deduced from the setup; that is, path-loss, antenna array design,  location of dominating interferers, receiver noise, etc. We therefore assume that $\alpha_{\mathrm{PEACH}}$ can be computed with low complexity in this paper and refer to \cite{Sessler2005a} for general techniques for efficient computation of eigenvalues.

\subsection{Weighted PEACH Estimator}

Although the PEACH estimator \eqref{eq_PEACH_estimator} converges to the MMSE estimator as $L \rightarrow \infty$, it is generally not the best $L$-order polynomial estimator at finite $L$. More specifically, instead of multiplying each term in the sum with $\alpha$, we can assign different weights and optimize these for the specific order $L$.
In this way, we obtain the \emph{weighted PEACH estimator}
\begin{equation} \label{eq_weighted_PEACH_estimator}
\begin{split}
\vectorize(\WPEACH{\vect{H}}) = \vect{R} \widetilde{\vect{P}}^H
\sum_{l=0}^{L} w_l \alpha^{l+1} \big(\widetilde{\vect{P}} \vect{R} \widetilde{\vect{P}}^H + \vect{S}  \big)^{l}  \vectorize(\vect{Y})
\end{split}
\end{equation}
where $\vect{w} = [w_0,\ldots,w_L]^T$ are scalar weighting coefficients.\footnote{W-PEACH is obtained by expanding each $(\vect{I} - \alpha (\widetilde{\vect{P}} \vect{R} \widetilde{\vect{P}}^H + \vect{S} ) )^{l}$ as a binomial series, collecting terms, and replacing constant factors with weights.} Observe that the $\alpha$-parameter is now redundant and can be set to one. For numerical reasons, it might still be good to select
\begin{equation}
\alpha_{\mathrm{W}\textrm{-}\mathrm{PEACH}} \leq \frac{1}{\max_n \lambda_n(\widetilde{\vect{P}} \vect{R} \widetilde{\vect{P}}^H + \vect{S})}
\end{equation}
since this prevent the eigenvalues of $\alpha^{l+1} \big(\widetilde{\vect{P}} \vect{R} \widetilde{\vect{P}}^H + \vect{S}  \big)^{l}$ to grow unboundedly as $l$ becomes large. This will simplify the implementation of the following theorem, which finds the weights that minimize the MSE.

\begin{theorem} \label{theorem:MSE-minimizing-weights}
The MSE $\mathbb{E}\{ \| \vect{H} - \WPEACH{\vect{H}} \|_F^2 \}$ is minimized by
\begin{equation} \label{eq_optimal_coefficients}
\vect{w}_{\mathrm{opt}} = [w_0^{\mathrm{opt}} \, \ldots \, w_L^{\mathrm{opt}}]^T = \vect{A}^{-1} \vect{b}
\end{equation}
where the $ij$th element of $\vect{A} \in \mathbb{C}^{L+1 \times L+1}$ and the $i$th element of $\vect{b} \in \mathbb{C}^{L+1}$ are
\begin{equation} \label{eq_A_b_matrices}
\begin{split}
[\vect{A}]_{ij} &= \alpha^{i+j}  \tr\left(  \vect{R} \widetilde{\vect{P}}^H (\widetilde{\vect{P}} \vect{R} \widetilde{\vect{P}}^H + \vect{S})^{i+j-1}     \widetilde{\vect{P}} \vect{R}  \right), \\
[\vect{b}]_i &= \alpha^{i} \tr\left( \vect{R} \widetilde{\vect{P}}^H (\widetilde{\vect{P}} \vect{R} \widetilde{\vect{P}}^H + \vect{S})^{i-1}  \widetilde{\vect{P}} \vect{R} \right).
\end{split}
\end{equation}
The resulting MSE of the W-PEACH estimator is
\begin{equation}
\tr(\vect{R}) + \vect{w}_{\mathrm{opt}}^H \vect{A} \vect{w}_{\mathrm{opt}}  - \vect{b}^H \vect{w}_{\mathrm{opt}}  - \vect{w}_{\mathrm{opt}}^H \vect{b}.
\end{equation}
\end{theorem}
\begin{IEEEproof}
Follows from differentiation of the MSE.
\end{IEEEproof}

Although Theorem \ref{theorem:MSE-minimizing-weights} provides the optimal weights, the computational complexity is $\mathcal{O}(M^3)$ since it involves pure matrix multiplications of the form $\vect{Z}^{i}$. This means that computing the optimal weights for the W-PEACH estimator has the same asymptotic complexity scaling as computing the original MMSE estimator. To benefit from the weighting we thus need to find a low-complexity approach to compute the weights.

\subsection{Low-Complexity Weights}

Next, we propose a low-complexity algorithm to compute weights for the W-PEACH estimator. We will exploit that
\begin{equation}
(\widetilde{\vect{P}} \vect{R} \widetilde{\vect{P}}^H + \vect{S}) = \mathbb{E}\{ \vectorize(\vect{Y}) \vectorize(\vect{Y})^H \} = \lim_{T \rightarrow \infty} \frac{1}{T} \sum_{t=1}^{T} \vect{y}_t \vect{y}_t^H,
\end{equation}
where $\vect{y}_t = \vectorize(\vect{Y})$ denotes the received signal at estimation time instant $t$. This means that $(\widetilde{\vect{P}} \vect{R} \widetilde{\vect{P}}^H + \vect{S})$ is closely approximated by the sample covariance matrix $\frac{1}{T} \sum_{t=1}^{T} \vect{y}_t \vect{y}_t^H$ if the number of samples $T$ is large. Although one generally needs $T \gg B N_r$ to get a good approximation, we can get away with much smaller $T$ since we will only compute traces.

For any fixed $T \geq 1$ and $i\geq 1$, we now observe that
\begin{align} \label{eq:low-complex1}
&\tr\left(  \vect{R} \widetilde{\vect{P}}^H (\widetilde{\vect{P}} \vect{R} \widetilde{\vect{P}}^H + \vect{S})^{i}     \widetilde{\vect{P}} \vect{R}  \right) \\
&\approx \tr\left(  \vect{R} \widetilde{\vect{P}}^H (\widetilde{\vect{P}} \vect{R} \widetilde{\vect{P}}^H + \vect{S})^{i-1} \left(\frac{1}{T} \sum_{t=1}^{T}  \vect{y}_t \vect{y}_t^H \right)    \widetilde{\vect{P}} \vect{R}  \right) \\
&= \frac{1}{T} \sum_{t=1}^{T} \vect{y}_t^H \left( \widetilde{\vect{P}} \vect{R}^2 \widetilde{\vect{P}}^H (\widetilde{\vect{P}} \vect{R} \widetilde{\vect{P}}^H + \vect{S})^{i-1} \right) \vect{y}_t. \label{eq:low-complex3}
\end{align}

Since the elements of $\vect{A}$ and $\vect{b}$ in \eqref{eq_A_b_matrices} are of the form in \eqref{eq:low-complex1}, we can approximate each element using \eqref{eq:low-complex3}.\footnote{Note that $b_0 = \tr( \widetilde{\vect{P}} \vect{R}^2 \widetilde{\vect{P}}^H )$ needs to be treated differently since there is no $(\widetilde{\vect{P}} \vect{R} \widetilde{\vect{P}}^H + \vect{S})$ term. For some predefined vectors $\vect{v}_i \! \sim \! \mathcal{CN}(\vect{0},\vect{I})$,  Algorithm \ref{algorithm_low-complexity} uses the approximation $\tr( \widetilde{\vect{P}} \vect{R}^2 \widetilde{\vect{P}}^H ) \approx \frac{\alpha}{T} \sum_{i=1}^{T} \vect{v}_i^H \widetilde{\vect{P}} \vect{R}^2 \widetilde{\vect{P}}^H \vect{v}_i$.} By computing/updating these approximations over a sliding time window of length $T$, we obtain Algorithm \ref{algorithm_low-complexity}. At any time instant $t$, this algorithm computes approximations of $\vect{A},\vect{b}$, denoted by $\widetilde{\vect{A}}_{t},\tilde{\vect{b}}_{t}$, by using the received signals $\vect{y}_t,\ldots,\vect{y}_{t-T+1}$.
These are used to compute approximate weights $\vect{w}_{\mathrm{approx},t}$. To reduce the amount of computations, $\widetilde{\vect{A}}_{t},\tilde{\vect{b}}_{t}$ are obtained from $\widetilde{\vect{A}}_{t-1},\tilde{\vect{b}}_{t-1}$ by adding one term per element based on the current received signal $\vect{y}_{t}$ and removing the impact of the old received signal $\vect{y}_{t-T}$ (which is now outside the time window). The algorithm can be initialized in any way; for example, by accumulating $T$ received signals to fill the time window.

\begin{algorithm}[t]
  \KwIn{Polynomial order $L$ and time window $T$\;}
  \KwIn{Current time $t$\;}
  \KwIn{New and old received signals $\vect{y}_t,\vect{y}_{t-T}$\;}
  \KwIn{Approximations $\widetilde{\vect{A}}_{t-1},\tilde{\vect{b}}_{t-1}$ at previous time $t\!-\!1$\;}
  Set $[\widetilde{\vect{A}}_{t}]_{ij} = [\widetilde{\vect{A}}_{t-1}]_{ij} $ \begin{displaymath}
  \begin{split}
  &+ \frac{ \alpha^{i+j}}{T} \vect{y}_t^H \left( \widetilde{\vect{P}} \vect{R}^2 \widetilde{\vect{P}}^H (\widetilde{\vect{P}} \vect{R} \widetilde{\vect{P}}^H + \vect{S})^{i+j-2} \right) \vect{y}_t \\
  &- \frac{ \alpha^{i+j}}{T} \vect{y}_{t-T}^H \left( \widetilde{\vect{P}} \vect{R}^2 \widetilde{\vect{P}}^H (\widetilde{\vect{P}} \vect{R} \widetilde{\vect{P}}^H + \vect{S})^{i+j-2} \right) \vect{y}_{t-T} \,\,\, \forall i,j\;
  \end{split}
  \end{displaymath}

  Set $[\tilde{\vect{b}}_{t}]_i = [\tilde{\vect{b}}_{t-1}]_i$ \begin{displaymath}
  \begin{split}
  & + \frac{ \alpha^{i}}{T} \vect{y}_t^H \left( \widetilde{\vect{P}} \vect{R}^2 \widetilde{\vect{P}}^H (\widetilde{\vect{P}} \vect{R} \widetilde{\vect{P}}^H + \vect{S})^{i-2} \right) \vect{y}_t \\& - \frac{ \alpha^{i}}{T} \vect{y}_{t-T}^H \left( \widetilde{\vect{P}} \vect{R}^2 \widetilde{\vect{P}}^H (\widetilde{\vect{P}} \vect{R} \widetilde{\vect{P}}^H + \vect{S})^{i-2} \right) \vect{y}_{t-T} \,\,\,\, \forall i \geq 2\;
     \end{split}
  \end{displaymath}

  Set $[\tilde{\vect{b}}_{t}]_1 = \frac{\alpha}{T} \sum_{i=1}^{T} \vect{v}_i^H \widetilde{\vect{P}} \vect{R}^2 \widetilde{\vect{P}}^H \vect{v}_i$ for $\vect{v}_i \! \sim \! \mathcal{CN}(\vect{0},\vect{I})$\;

  Compute $\vect{w}_{\mathrm{approx},t} = \widetilde{\vect{A}}_{t}^{-1} \tilde{\vect{b}}_{t}$\;
  \KwOut{Approximate weights $\vect{w}_{\mathrm{approx},t}$ at time $t$\;}
\caption{Low-complexity weights for W-PEACH} \label{algorithm_low-complexity}
\end{algorithm}

The complexity of computing the elements in $\widetilde{\vect{A}}_{t},\tilde{\vect{b}}_{t}$ is $\mathcal{O}(L^2 M^2)$ \emph{per time instant} since there are $(L+1)(L+1)$ elements in $\widetilde{\vect{A}}_{t}$ and $L+1$ elements in $\tilde{\vect{b}}_{t}$ to be computed, which results in $(L+2)(L+1)$ elements in total, and we need to compute  a series of multiplications between vectors and matrices. Furthermore, $\vect{w}_{\mathrm{approx},t}$ is obtained by solving an $L$-dimensional system of equations, which has complexity $\mathcal{O}(L^3)$. To summarize, the W-PEACH estimator along with Algorithm \ref{algorithm_low-complexity} has a computational complexity of $\mathcal{O}(L^2 M^2 + L M^2 + L^3) = \mathcal{O}(L^2 M^2 + L^3)$, which is smaller than $\mathcal{O}(M^3)$ of the MMSE estimator for $L < \sqrt{M} = \sqrt{N_t N_r}$.

One additional feature of Algorithm \ref{algorithm_low-complexity} is that it can easily be extended to practical scenarios where also the true covariance matrices $\vect{R}$ and $(\widetilde{\vect{P}} \vect{R} \widetilde{\vect{P}}^H + \vect{S})$ are approximated by sample covariance matrices. This would enable adaptive tracking of the slow variations in channel and disturbance statistics that appear in practice. We leave this extension for future work.

\subsection{Summary of Computational Complexity}

The complexity of the conventional estimators and proposed PEACH estimators are summarized as follows:

\vskip+2mm

\begin{center}
     \begin{tabular}{ | c | c |}
     \hline
     Channel Estimators & Computational Complexity \\ \hline
     MMSE and MVU & $\mathcal{O}(N_t^3 N_r^3)$ \\ \hline
     PEACH & $\mathcal{O}(L N_t^2 N_r^2)$ \\ \hline
     W-PEACH & $\mathcal{O}(L^2 N_t^2 N_r^2 + L^3)$ \\ \hline
     \end{tabular}
 \end{center}

\vskip+2mm

We note that the cubic complexity scaling in $N_t N_r$ for the conventional MMSE and MVU estimators have been reduced to squared complexity for the proposed PEACH estimators. The order $L$ of the polynomial expansion has a clear impact on the complexity, but recall that it generally needs not scale with $N_t N_r$ \cite{Honig2001a}. In the next section, we will illustrate that small values on $L$ yields good performance.

\section{Numerical Evaluation}

In this section, we illustrate the performance of the proposed PEACH and W-PEACH estimators. The analysis so far has been generic with respect to the disturbance covariance matrix $\vect{S}$. Here, we consider two scenarios: noise-limited and cellular networks with pilot contamination. We describe the latter scenario  in more detail since it is one of the main challenges in the development of large-scale MIMO systems \cite{Rusek2013a}.

\subsection{Noise-Limited Scenario}
\label{subsection:noise-limited}

A commonly studied scenario is when there is only uncorrelated receiver noise; thus $\vect{S} = \sigma^2 \vect{I}$ where $\sigma^2$ is the variance.

\subsection{Pilot Contamination Scenario}
\label{subsection:pilot-contamination}

A scenario that has received much attention in the large-scale MIMO literature is when there is
disturbance from simultaneous reuse of pilot signals in neighboring cells \cite{Jose2011b,Hoydis2013a,Rusek2013a,Yin2013a,Mueller2013a}. Such reuse is often necessary due to the finite channel coherence time (i.e., the time that a channel estimate can be considered accurate), but leads to a special form of interference called pilot contamination. It can be modeled as\footnote{Cell $i$ can use an arbitrary pilot $\vect{P}_i$, but only pilots with overlapping span (i.e., $\vect{P}_i \vect{P}^H \neq \vect{0}$) will cause interference. Therefore, the case of a common reused pilot $\vect{P}_i =\vect{P}$ is of main interest, while extensions are straightforward.}
\begin{equation}
\vect{N} = \sum_{i \in \mathcal{I}} \vect{H}_i \vect{P} + \widetilde{\vect{N}}
\end{equation}
where $\mathcal{I}$ is the set of interfering cells, $\vect{H}_i$ is the channel from the $i$th interfering cell to the receiver in the cell under study, and $\vectorize(\widetilde{\vect{N}}) \sim \mathcal{CN}(\vect{0},\sigma^2 \vect{I})$ is uncorrelated receiver noise. If $\vect{H}_i$ is Rayleigh fading with $\vectorize(\vect{H}_i) \sim \mathcal{CN}(\vect{0},\boldsymbol{\Sigma}_i)$, then
\begin{equation} \label{eq:pilot-contamination-covariance}
\vect{S} =  \sum_{i \in \mathcal{I}} \widetilde{\vect{P}} \boldsymbol{\Sigma}_i \widetilde{\vect{P}}^H + \sigma^2 \vect{I}.
\end{equation}
Note that only the sum covariance matrix $\sum_{i \in \mathcal{I}}  \boldsymbol{\Sigma}_i$ needs to be known when computing the proposed PEACH estimators.

When \eqref{eq:pilot-contamination-covariance} is substituted into the PEACH and W-PEACH estimator expressions in \eqref{eq_PEACH_estimator} and \eqref{eq_weighted_PEACH_estimator} we get contaminated terms of the form $\vect{R} \widetilde{\vect{P}}^H \widetilde{\vect{P}} \boldsymbol{\Sigma}_i \widetilde{\vect{P}}^H$. These terms are small if $\vect{R}$ and $\boldsymbol{\Sigma}_i $ have very different span, or if $\tr(\boldsymbol{\Sigma}_i)$ is weak altogether---this is easily observed if $\widetilde{\vect{P}}^H \widetilde{\vect{P}}$ is a scaled identity matrix. Similar observations were recently made in the capacity analysis of [3] and when developing a pilot allocation algorithm in [5]. Under certain conditions, the subspaces of the useful channel and pilot contamination can be made orthogonal by coordinated pilot allocation across cells [5] or by exploiting both received pilot and data signals for channel estimation as in \cite{Mueller2013a}.

\subsection{Numerical Examples}
\label{subsection:examples}

To evaluate the performance of our proposed estimators, we consider a large-scale MIMO system with $N_r = 100$ and $N_t = 10$ antennas in addition to the pilot length $B = 10$. We follow the Kronecker model \cite{Shiu2000a} to describe correlation among antennas of the desired and disturbance MIMO channels. In this model, the covariance matrix of a MIMO channel is modeled as $\vect{R} = \vect{R_t} \otimes \vect{R_r}$ where $\vect{R_t} \in \mathbb{C}^{N_t \times N_t}$ and $\vect{R_r} \in \mathbb{C}^{N_r \times N_r}$ are the spatial covariance matrices at the transmitter and receiver sides, respectively. Following the same modeling, we have $\boldsymbol{\Sigma}_i  = \boldsymbol{\Sigma}_{\vect{t}_i}  \otimes \boldsymbol{\Sigma}_{\vect{r}_i}$ for $i \in \mathcal{I}$.

\begin{figure}
   \begin{center}
   \includegraphics[width=\columnwidth]{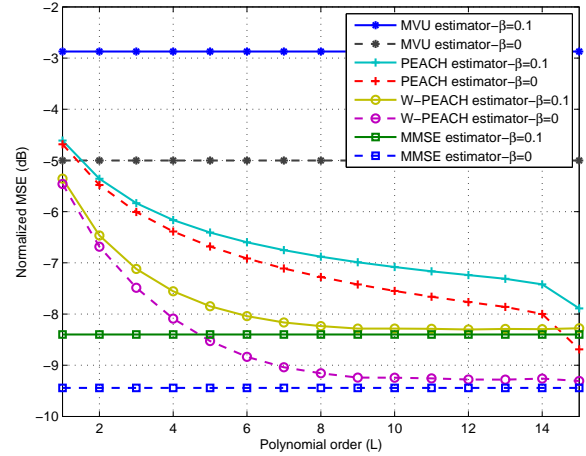}\\ \vskip-3mm
 \caption{MSE comparison of different estimators for different polynomial orders when $\beta \in \{ 0, \, 0.1\}$, which represents noise-limited and pilot contaminated scenarios.}
   \label{fig:Fig1}
   \end{center} \vskip-6mm
\end{figure}

To generate covariance matrices, we use the exponential model \cite{Loyka2001a}. Without loss of generality, all the covariance matrices will have diagonal elements equal to one which results in $\tr( \vect{R} ) = \tr( \boldsymbol{\Sigma}_i ) = N_tN_r$. We assume that there are two dominating interfering cells, $i=1,2$, for which the covariance matrices are weakened by the factor $0 \leq \beta_i < 1$, i.e., $\beta_i  \boldsymbol{\Sigma}_i $. This factor represents how severe the pilot contamination part is: $\beta_i=0$ represents the noise-limited case, while $\beta_i=1$ represents the case when the useful channel and the $i$th interfering channel are equally strong. We define the normalized pilot signal-to-noise ratio (SNR) as $\gamma = \frac{ \mathcal{P}_T }{ \sigma^2 }$
where $ \mathcal{P}_T = \frac{1}{N_t} \tr( \vect{PP}^H)$ is the average pilot power.

We use the normalized MSE, defined as $\frac{ \textrm{MSE}}{ \tr (\vect{R}) }$, as performance measure.
In all the figures, we compare the performance of the proposed estimators with the conventional MMSE and MVU estimators. The pilot matrix is $ \vect{P} = \sqrt{\mathcal{P}_T} \vect{I}$.

In Fig.~\ref{fig:Fig1}, MSE has been plotted as a function of the polynomial order $L$. The noise-limited scenario is given by $\beta = 0$, while $\beta=0.1$ (we assume that $\beta_1=\beta_2=\beta$) represents the scenario where the two interfering cells have interfering channels which are $10$ dB weaker than the desired channels. The SNR is $\gamma = 5$ dB. As can be seen from Fig.~\ref{fig:Fig1}, both PEACH and W-PEACH detract by increasing $L$ and outperform MVU for $L\geq2$.
Interestingly, W-PEACH approaches the MSE-values of the MMSE estimator very quickly, while PEACH needs a higher $L$ than W-PEACH to get close to the MMSE curves.

\begin{figure}
   \begin{center}
   \includegraphics[width=\columnwidth]{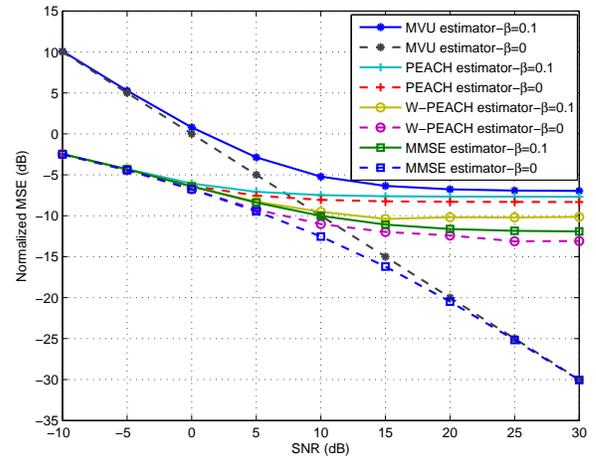}\\ \vskip-3mm
 \caption{MSE comparison of different estimators for different values of the SNR $\gamma$ when $\beta \in \{ 0, \, 0.1\}$, which represents noise-limited and pilot contaminated scenarios.}
   \label{fig:Fig2}
   \end{center} \vskip-7mm
\end{figure}

In Fig.~\ref{fig:Fig2}, we compare different estimators with or without additional interference from pilot contamination. We consider a fixed $L=10$ and vary the SNR $\gamma$. As expected, the MSE of the MMSE and MVU estimators decay steeply to zero when the $\gamma$ increases in the noise-limited scenario, while the MSE saturates to a non-zero error floor under pilot contamination. The MSE floor when $\beta \neq 0$ is due to the fact that the normalized signal-to-interference-and-noise ratio (SINR) becomes almost constant as $\gamma$ increases. To show this, we define the normalized SINR as
\begin{equation} \label{eq:SINR-expression}
\frac{\gamma}{1 + K \beta \gamma}
\end{equation}
where $K$ is the number of interferers. As $\gamma$ increases, the SINR in \eqref{eq:SINR-expression} approaches $\frac{1}{K \beta}>0$, then the MSE values, which are different functions of the SINR for different estimators, approach some non-zero limits.

We observe from Fig.~\ref{fig:Fig2} that pilot contamination does not make significant impact on the PEACH and W-PEACH estimators;
in fact, pilot contamination is beneficial in the sense that it reduces the gap to the optimal MMSE estimator. This remarkable result is explained as follows. For any fixed $L$, PEACH and W-PEACH will converge to a non-zero MSE when $\gamma$ increases, due to the bias generated by the approximation error. Since this also happens for the MMSE and MVU estimators under pilot contamination, the relative loss of using the proposed low-complexity estimators will be smaller. Consequently, we can reduce $L$ as $\beta$ increases and still achieve near-optimal performance.

\begin{figure}
   \begin{center}
   \includegraphics[width=\columnwidth]{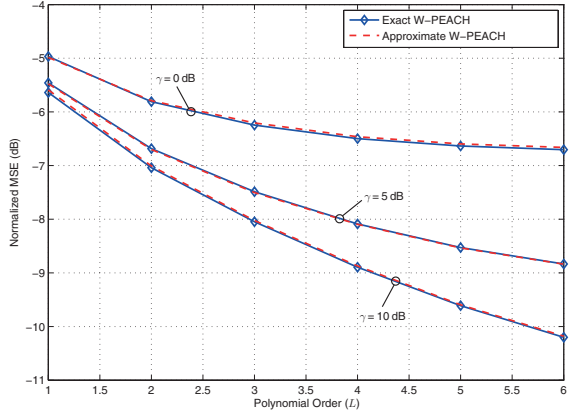}\\ \vskip-3mm
 \caption{Comparison of W-PEACH estimator and Approximate W-PEACH estimator in a noise-limited scenario ($\beta=0$) for different SNR $\gamma$ values.}
   \label{fig:Fig3}
   \end{center}  \vskip-7mm
\end{figure}

Finally, we focus on the low-complexity approach in Algorithm \ref{algorithm_low-complexity} for finding the weights.
Fig.~\ref{fig:Fig3} considers a noise-limited scenario and a time window of length $T=100$. Although $T \ll B N_r$, we observe that the approximate W-PEACH estimator which exploits the approximate weights from Algorithm \ref{algorithm_low-complexity} gives almost identical performance as W-PEACH with optimal weights computed according to Theorem \ref{theorem:MSE-minimizing-weights}. This confirms that the W-PEACH estimator is indeed a low-complexity channel estimator suitable for large-scale MIMO systems.

\section{Conclusions}

Large-scale MIMO techniques offer a high spatial resolution that can drastically increase the spectral and/or energy efficiency of wireless systems. Acquiring accurate CSI is a key to achieve these potential improvements in practice. Since the conventional pilot-based MMSE and MVU channel estimators have a computational complexity unsuitable for scenarios with large numbers of antennas, we proposed a set of low-complexity PEACH estimators. These are based on approximating the inversion of covariance matrices in the MMSE estimator by an $L$-order matrix polynomial. The proposed estimators converge to the MMSE estimator as $L$ grows large, but by deriving the optimal coefficients for the polynomial we can obtain near-optimal MSE performance at small values on $L$.
In practice, the order $L$ can be selected to balance between complexity and MSE performance. Numerical results are given for noise-limited scenarios as well as under pilot contamination from pilot reuse in adjacent systems. Although pilot contamination can create an MSE floor, it is actually beneficial to the proposed estimators in the sense that smaller $L$ can be used to achieve good performance.

\bibliographystyle{IEEEtran}
\bibliography{IEEEabrv,refs}

\begin{thebibliography}{10}
\providecommand{\url}[1]{#1}
\csname url@samestyle\endcsname
\providecommand{\newblock}{\relax}
\providecommand{\bibinfo}[2]{#2}
\providecommand{\BIBentrySTDinterwordspacing}{\spaceskip=0pt\relax}
\providecommand{\BIBentryALTinterwordstretchfactor}{4}
\providecommand{\BIBentryALTinterwordspacing}{\spaceskip=\fontdimen2\font plus
\BIBentryALTinterwordstretchfactor\fontdimen3\font minus
  \fontdimen4\font\relax}
\providecommand{\BIBforeignlanguage}[2]{{%
\expandafter\ifx\csname l@#1\endcsname\relax
\typeout{** WARNING: IEEEtran.bst: No hyphenation pattern has been}%
\typeout{** loaded for the language `#1'. Using the pattern for}%
\typeout{** the default language instead.}%
\else
\language=\csname l@#1\endcsname
\fi
#2}}
\providecommand{\BIBdecl}{\relax}
\BIBdecl

\bibitem{Holma2012a}
H.~Holma and A.~Toskala, \emph{{LTE Advanced: 3GPP Solution for IMT-Advanced}},
  1st~ed.\hskip 1em plus 0.5em minus 0.4em\relax Wiley, 2012.

\bibitem{Jose2011b}
J.~Jose, A.~Ashikhmin, T.~Marzetta, and S.~Vishwanath, ``Pilot contamination
  and precoding in multi-cell {TDD} systems,'' \emph{{IEEE} Trans. Commun.},
  vol.~10, no.~8, pp. 2640--2651, 2011.

\bibitem{Hoydis2013a}
J.~Hoydis, S.~ten Brink, and M.~Debbah, ``Massive {MIMO} in the {UL/DL} of
  cellular networks: How many antennas do we need?'' \emph{{IEEE} J. Sel. Areas
  Commun.}, vol.~31, no.~2, pp. 160--171, 2013.

\bibitem{Rusek2013a}
F.~Rusek, D.~Persson, B.~Lau, E.~Larsson, T.~Marzetta, O.~Edfors, and
  F.~Tufvesson, ``Scaling up {MIMO}: Opportunities and challenges with very
  large arrays,'' \emph{{IEEE} Signal Process. Mag.}, vol.~30, no.~1, pp.
  40--60, 2013.

\bibitem{Yin2013a}
H.~Yin, D.~Gesbert, M.~Filippou, and Y.~Liu, ``A coordinated approach to
  channel estimation in large-scale multiple-antenna systems,'' \emph{{IEEE} J.
  Sel. Areas Commun.}, vol.~31, no.~2, pp. 264--273, 2013.

\bibitem{Mueller2013a}
R.~M\"{u}ller, M.~Vehkaper\"{a}, and L.~Cottatellucci, ``Blind pilot
  decontamination,'' in \emph{Proc.~ITG Workshop on Smart Antennas (WSA)},
  2013.

\bibitem{Kay1993a}
S.~Kay, \emph{Fundamentals of Statistical Signal Processing: Estimation
  Theory}.\hskip 1em plus 0.5em minus 0.4em\relax Prentice Hall, 1993.

\bibitem{Kotecha2004a}
J.~Kotecha and A.~Sayeed, ``Transmit signal design for optimal estimation of
  correlated {MIMO} channels,'' \emph{{IEEE} Trans. Signal Process.}, vol.~52,
  no.~2, pp. 546--557, 2004.

\bibitem{Liu2007a}
Y.~Liu, T.~Wong, and W.~Hager, ``Training signal design for estimation of
  correlated {MIMO} channels with colored interference,'' \emph{{IEEE} Trans.
  Signal Process.}, vol.~55, no.~4, pp. 1486--1497, 2007.

\bibitem{Bjornson2010a}
E.~Bj{\"{o}}rnson and B.~Ottersten, ``A framework for training-based estimation
  in arbitrarily correlated {Rician} {MIMO} channels with {Rician}
  disturbance,'' \emph{{IEEE} Trans. Signal Process.}, vol.~58, no.~3, pp.
  1807--1820, 2010.

\bibitem{Moshavi1996a}
S.~Moshavi, E.~Kanterakis, and D.~Schilling, ``Multistage linear receivers for
  {DS-CDMA} systems,'' \emph{Int. J. Wireless Information Networks}, vol.~3,
  no.~1, pp. 1--17, 1996.

\bibitem{Honig2001a}
M.~Honig and W.~Xiao, ``Performance of reduced-rank linear interference
  suppression,'' \emph{{IEEE} Trans. Inf. Theory}, vol.~47, no.~5, pp.
  1928--1946, 2001.

\bibitem{Sessler2005a}
G.~Sessler and F.~Jondral, ``Low complexity polynomial expansion multiuser
  detector for {CDMA} systems,'' \emph{{IEEE} Trans. Veh. Technol.}, vol.~54,
  no.~4, pp. 1379--1391, 2005.

\bibitem{Hoydis2011d}
J.~Hoydis, M.~Debbah, and M.~Kobayashi, ``Asymptotic moments for interference
  mitigation in correlated fading channels,'' in \emph{Proc.~IEEE ISIT}, 2011.

\bibitem{Williams2012a}
V.~Williams, ``Multiplying matrices faster than {Coppersmith-Winograd},'' in
  \emph{Proc. STOC}, 2012, pp. 887--898.

\bibitem{Shiu2000a}
D.-S. Shiu, G.~Foschini, M.~Gans, and J.~Kahn, ``Fading correlation and its
  effect on the capacity of multielement antenna systems,'' \emph{{IEEE} Trans.
  Commun.}, vol.~48, no.~3, pp. 502--513, 2000.

\bibitem{Loyka2001a}
S.~Loyka, ``Channel capacity of {MIMO} architecture using the exponential
  correlation matrix,'' \emph{{IEEE} Commun. Lett.}, vol.~5, no.~9, pp.
  369--371, 2001.

\end{thebibliography}

\end{document}